\documentclass[pdftex,twocolumn,epjc3]{svjour3}

\RequirePackage[T1]{fontenc}
\RequirePackage{graphicx}
\RequirePackage{mathptmx}      % use Times fonts if available on your TeX system
\RequirePackage{flushend}
\RequirePackage[numbers,sort&compress]{natbib}
\RequirePackage[colorlinks,citecolor=blue,urlcolor=blue,linkcolor=blue]{hyperref}

\usepackage{color}
\definecolor{darkgreen}{rgb}{0,0.5,0}
\definecolor{darkblue}{rgb}{0,0,0.7}
\definecolor{darkred}{rgb}{0.5,0,0.0}
\definecolor{darkorange}{rgb}{0.8,0.4,0.0}

\usepackage{amsmath}
\usepackage{amssymb}
\usepackage{xspace}

\journalname{Eur. Phys. J. C}

\newcommand{\zcut}{z_{\text{cut}}}
\newcommand{\SD}{Soft Drop\xspace}
\newcommand{\Lhadr}{\ensuremath{\Lambda_{\text{hadr}}}}
\newcommand{\zsd}{z_{\text{SD}}}
\newcommand{\Lsd}{L_{\text{SD}}}

\newcommand{\herwig}{\textsf{Herwig}}
\newcommand{\pythia}{\textsf{Pythia}}

\begin{document}

\title{The jet mass distribution after \SD}
\author{Simone Marzani\thanksref{e1,addr1}
        \and
        Lais Schunk\thanksref{e2,addr2}
        \and
        Gregory Soyez\thanksref{e3,addr3}
}

\thankstext{e1}{e-mail: simone.marzani@ge.infn.it}
\thankstext{e2}{e-mail: lais.sarem.schunk@desy.de}
\thankstext{e3}{e-mail: gregory.soyez@ipht.fr}

\institute{Dipartimento di Fisica, Universit\`a di Genova and INFN, Sezione di Genova, Via Dodecaneso 33, 16146, Italy\label{addr1}
          \and
          Deutsches Elektronen-Synchroton DESY, Notkestra\ss e 85, 22607 Hamburg, Germany\label{addr2}
          \and
          IPhT, CEA Saclay, CNRS UMR 3681, F-91191 Gif-Sur-Yvette, France\label{addr3}
}

\date{December 2017}

%\institute{teste}

\maketitle

\begin{abstract}
  We present a first-principle computation of the mass distribution of
  jets which have undergone the grooming procedure known as \SD.
  This calculation includes the resummation of the large logarithms of
  the jet mass over its transverse momentum, up to next-to-logarithmic
  accuracy, matched to exact fixed-order results at next-to-leading
  order.
  We also include non-perturbative corrections obtained from
  Monte-Carlo simulations and discuss analytic expressions for
  hadronisation and Underlying Event effects.
\end{abstract}

%%%%%%%%%%%%%%%%%%%%%%%%%%%%%%%%%%%%%%%%%%%%%%%%%%%%%%%%%%%%%%%%%%%%%%%%%%
%\section{Introduction}\label{sec:intro}
\paragraph{Introduction.}
The study of jets at the Large Hadron Collider (LHC) has recently
taken a new turn with new substructure observables~\cite{mMDT,SD}
amenable to precise theory
calculations~\cite{flsy-letter,flsy,our_mmdt}, including genuine
theory uncertainty bands, and corresponding experimental measurements
from both the CMS~\cite{cms_mmdt} and ATLAS~\cite{atlas-sd}
collaborations.
The substructure techniques we concentrate on are usually referred to
as {\it grooming} and they aim to reduce sensitivity to
non-perturba\-tive corrections and pileup.

A first series of studies has focused on the jet mass after applying
the (modified) MassDrop~Tagger (mMDT)~\cite{MDT,mMDT} in dijet events,
as measured by the CMS collaboration~\cite{cms_mmdt}.
On the theory side, the description of this observable requires to
match a resummed calculation, important in the small-mass region, to
fixed-order results, which are important for large masses. The former
are obtained analytically, including to all orders terms enhanced by
the large logarithms of $p_t^2/m^2$ with $p_t$ the jet transverse
momentum and $m$ the (groomed) jet mass.
The latter is obtained from fixed-order Monte-Carlo simulations.
To date, two theory calculations are available: a SCET-based
next-to-leading logarithmic (NLL) resummation in the small $\zcut$
limit, matched to leading order (LO) results~\cite{flsy}, and our
previous study matching a leading logarithmic resummation, including
finite (but small) $\zcut$ effects, to next-to-leading order
results~\cite{our_mmdt}.
Comparing both predictions, we see a small NLL effect at small mass
and non-negligible NLO corrections at large mass.

The goal of the present letter is to extend our mMDT study from
Ref.~\cite{our_mmdt} to the case of \SD~\cite{SD}, i.e.\ allowing for
a non-zero value of the angular exponent $\beta$.
When $\beta \neq 0$, the logarithmic counting differs from the mMDT
case, essentially because \SD retains soft-collinear radiation, which
is always groomed away by mMDT.
In this case, the SCET-based calculation from Ref.~\cite{flsy} reaches
NNLL accuracy and it is matched, in the dijet case, to LO fixed-order
results. Here, we present the results of a NLL resummation matched to
NLO fixed-order accuracy.

After a brief review of the \SD procedure, we will present our results
first in the resummation region, then match\-ed to fixed-order. We
then provide an analytic estimate of non-perturbative corrections,
extending to the \SD case the analytic results obtained in
Ref.~\cite{mMDT} for the mMDT. We conclude by providing and discussing
our final predictions, including the theory uncertainty bands.
These have already been compared to experimental data
in~\cite{atlas-sd}, where a good agreement was found, especially in
the perturbative region.

%%%%%%%%%%%%%%%%%%%%%%%%%%%%%%%%%%%%%%%%%%%%%%%%%%%%%%%%%%%%%%%%%%%%%%%%%%
%\section{\SD}\label{sec:softdrop}
\paragraph{\SD.}

For a given jet, the \SD procedure first re-clusters the constituents
of the jet with the Cambridge/Aachen algorithm~\cite{cam} into a
single jet $j$. Starting from $j$, it then applies the following
iterative procedure:
\begin{enumerate}
\item undo the last clustering step $j\to j_1,j_2$, with
  $p_{t1}>p_{t2}$.\label{sd-step1}
\item stop the procedure if the \SD condition is met:
\begin{equation}\label{eq:sd-cdt}
\frac{\text{min}(p_{t1},p_{t2})}{p_{t1}+p_{t2}}
   > \zcut \Big(\frac{\theta_{12}}{R}\Big)^\beta,
\end{equation}
where $\zcut$ and $\beta$ are free parameters,
$\theta_{12}^2=\Delta y_{12}^2+\Delta \phi_{12}^2$ and $R$ the
original jet radius.
\item otherwise, set $j=j_1$ and go back to~\ref{sd-step1}, or stop if
  $j_1$ has no further substructure.
\end{enumerate}

The limit $\beta\to 0$ corresponds to the mMDT.

%%%%%%%%%%%%%%%%%%%%%%%%%%%%%%%%%%%%%%%%%%%%%%%%%%%%%%%%%%%%%%%%%%%%%%%%%%
%\section{NLL resummation}\label{sec:resum}
\paragraph{NLL resummation.}

We consider the cumulative cross-section for the ratio $m^2/(p_tR)^2$ 
to be smaller than some value $\rho$, integrated over the
${\cal{O}}(\alpha_s^2)$ matrix element for the Born-level production
of 2 jets, in a given $p_t$ bin: 
\begin{equation}\label{eq:nll-resum}
\Sigma_{\text{NLL}}(\rho;p_{t1},p_{t2})=\int_{p_{t1}}^{p_{t2}} \!\!dp_t
\sum_i \frac{d\sigma_{\text{jet,LO}}^{(i)}}{dp_t} \frac{e^{-R_i(\rho)-\gamma_ER_i'(\rho)}}{\Gamma(1+R_i'(\rho))},
\end{equation}
where we have separated contributions from different flavour channels,
$R_i'$ is the derivative of $R_i$ wrt $\log(1/\rho)$ and the radiator
$R_i$ is given by
\begin{align}
&R_i(\rho) = \frac{C_i}{2\pi\alpha_s\beta_0^2}
  \bigg\{
    \Big[
      W(1-\lambda_B)-\frac{W(1-\lambda_c)}{1+\beta}-2W(1-\lambda_1)\nonumber\\
&\quad      +\frac{2+\beta}{1+\beta}W(1-\lambda_2)
    \Big]
-\frac{\alpha_s K}{2\pi}
\Big[\log(1-\lambda_B)-\frac{\log(1-\lambda_c)}{1+\beta}\nonumber\\
&\quad+\frac{2+\beta}{1+\beta}\log(1-\lambda_2) -2\log(1-\lambda_1)
  \Big] +\frac{\alpha_s \beta_1}{\beta_0}
\Big[V(1-\lambda_B)\nonumber\\
&\quad-\frac{V(1-\lambda_c)}{1+\beta}-2V(1-\lambda_1)
+\frac{2+\beta}{1+\beta}V(1-\lambda_2)  
  \Big]
\bigg\}\,,\label{eq:radiator}
\end{align}
where
\begin{align}
 \lambda_c = 2\alpha_s\beta_0\log(1/\zcut), && 
 \lambda_\rho = 2\alpha_s\beta_0\log(1/\rho), \\
 \lambda_1 = \frac{\lambda_\rho+\lambda_B}{2}, && 
 \lambda_2 = \frac{\lambda_c+(1+\beta)\lambda_\rho}{2+\beta},
\end{align}
and $\lambda_B=2\alpha_S\beta_0B_i$ appears due to hard-collinear
splittings, and $W(x)=x\log(x)$, $V(x)=\frac{1}{2}\log^2(x)+\log(x)$.

Note that $\alpha_s$ is calculated using the exact two-loop running
coupling, at the scale $p_tR$, and, in order to reach NLL accuracy, it is evaluated in
the CMW scheme~\cite{cmw}.
Furthermore, compared to the original results~\cite{SD}, the
hard-collinear contributions have been expressed as corrections to
double-logarithm arguments. In practice, this is equivalent to
replacing $P_i(z)\to (2C_i/z)\Theta(z<e^{B_i})$.  This introduces
unwanted NNLL terms but has the advantage to give well-defined and
positive resummed distributions which, in turn, makes the matching to
fix order easier.

To avoid any potential issue related to the Landau pole, appearing in
a region anyway dominated by hadronisation, we have frozen the
coupling at a scale $\mu_{\text{fr}}=1$~GeV. Corresponding expressions
can be found e.g. in Ref.~\cite{shapes-paper}.

%%%%%%%%%%%%%%%%%%%%%%%%%%%%%%%%%%%%%%%%%%%%%%%%%%%%%%%%%%%%%%%%%%%%%%%%%%
%\section{Matching to NLO}\label{sec:matching}
\paragraph{Matching to NLO.}

The \SD mass distributions for the dijet processes can be calculated
at fixed order at ${\cal{O}}(\alpha_s^4)$, i.e. up to NLO
accuracy. This is available for example using the
NLOJet++~\cite{nlojet} generator to simulate $2\to 3$ events at LO and
NLO. Jets are then clustered with the anti-$k_t$
algorithm~\cite{antikt} as implemented in
FastJet-3.2.2~\cite{fastjet}. In what follows, we have used the CT14
PDF set~\cite{CT14}.

NLO mass distributions need to be matched to our NLL resummed results.
For this, the LO jet mass distribution needs to be separated in flavour
channels, while the flavour separation of the NLO jet mass
distribution is instead subleading.
At ${\cal{O}}(\alpha_s^3)$ a jet has at most two constituent and the
only case where the flavour is ambiguous is when a jet is made of two
quarks (or a quark and an anti-quark of different flavours).
We (arbitrarily) treat this as a quark jet, an approximation which is
valid at our accuracy.
To keep the required flavour information in NLOJet++, we have used the
patch introduced in Ref.~\cite{bsz-shapes}.

To avoid artefacts at large mass, the endpoint of the resummed
calculation is matched to the endpoint of the perturbative
distribution by replacing
\begin{equation}
\log\Big(\frac{1}{\rho}\Big) \to 
\log\Big(\frac{1}{\rho}-\frac{1}{\rho_{\text{max},i}}+e^{-{B_q}}\Big)
\end{equation}
in the resummed results~\cite{match-endpoint}. The endpoints of the LO
and NLO distributions are found to be (see Appendix B of
Ref.~\cite{our_mmdt}) $\rho_{\text{max,LO}} \approx 0.279303$ and
$\rho_{\text{max,NLO}} \approx 0.44974$, for $R=0.8$.

\begin{figure*}
  \includegraphics[width=0.48\textwidth,page=1]{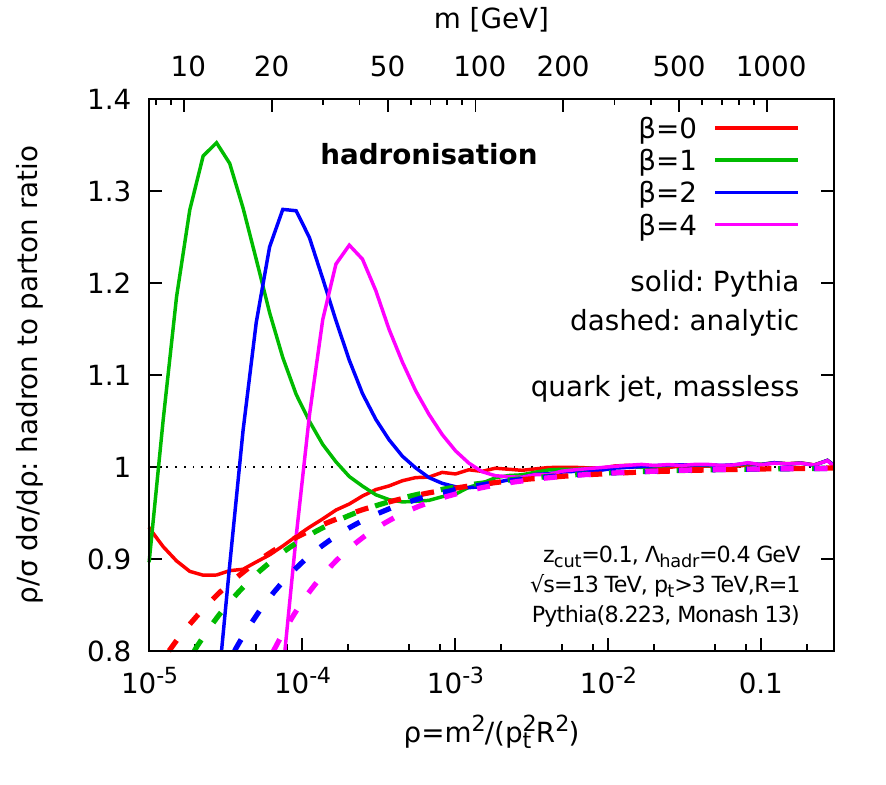}%
  \hfill%
  \includegraphics[width=0.48\textwidth,page=1]{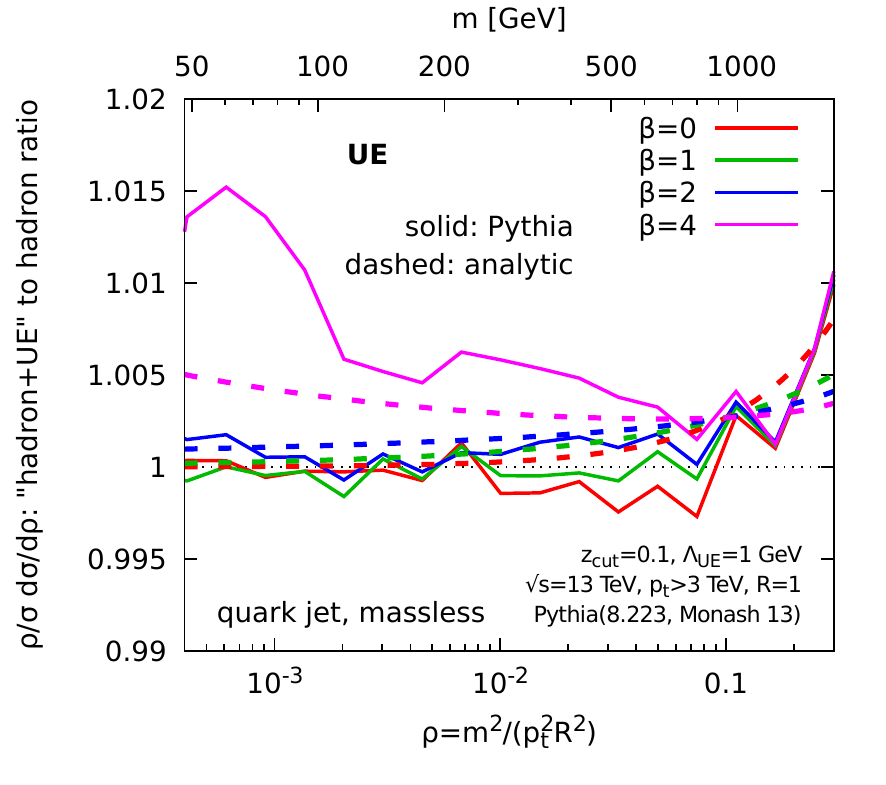}
  \caption{Comparison of our theoretical estimate of hadronisation
    corrections to what is implemented in a standard Monte-Carlo
    parton shower, for different values of the angular exponent
    $\beta$ , $\zcut =0.1$ and $R=0.8$. Quark jets are considered and hadron masses are
    neglected. Left: hadronisation corrections (i.e. ratio of hadron
    level to parton level) for $\Lhadr = 0.4$~GeV; right: Underlying Event corrections
    (i.e. ratio of distributions with and without UE) for $\Lambda_\text{UE}= 1$~GeV.}
\label{fig:np-ratio}
\end{figure*}

Finally, the matching between NLL and NLO results in each $p_t$ bin
can be done using log-R matching given by~\cite{bsz-shapes}
\begin{align}\label{eq:matched}
&\Sigma_{\text{NLL+NLO}}(\rho) =
\Bigg[ \sum_i
  \Sigma_{\text{NLL}}^{(i)}
  \exp\bigg(\frac{\Sigma_{\text{LO}}^{(i)}-\Sigma_{\text{NLL,LO}}^{(i)}}
                 {\sigma_{\text{jet,LO}}^{(i)}}
      \bigg)
\Bigg]\\
&\times \exp
  \Bigg(
    \frac{\bar\Sigma_{\text{NLO}}-\Sigma_{\text{NLL,NLO}}}
         {\sigma_{\text{jet,LO}}}
   -\sum_i\frac{(\Sigma_{\text{LO}}^{(i)})^2-(\Sigma_{\text{NLL,LO}}^{(i)})^2}{\sigma_{\text{jet,LO}}^{(i)}\sigma_{\text{jet,LO}}}\Bigg).\nonumber
\end{align}
In this expression, $\Sigma_{\text{NLL}}^{(i)}$ is given by
Eq.~(\ref{eq:nll-resum}), trivially split in flavour channels.
$\Sigma_{\text{NLL,LO}}^{(i)}$ and $\Sigma_{\text{NLL,NLO}}$ (summed
over flavour channels) are the expansion of
$\Sigma_{\text{NLL}}^{(i)}$ to LO (${\cal{O}}(\alpha_s^3)$) and NLO
(${\cal{O}}(\alpha_s^4)$), respectively. For the fixed-order part
\begin{align}
\Sigma_{\text{LO}}^{(i)}
  &=-\int_\rho^1 d\rho'\,\frac{d\sigma_{\text{mass,LO}}^{(i)}}{d\rho'}
    + \sigma_{\text{jet,NLO}}^{(i)},\\
\bar\Sigma_{\text{NLO}}
  &=-\int_\rho^1 d\rho'\,\frac{d\sigma_{\text{mass,NLO}}}{d\rho'},
\end{align}
where $d\sigma_{\text{mass,(N)LO}}/d\rho$ denotes the mass
distribution at (N)LO as obtained from NLOJet++ and
$\sigma_{\text{jet,(N)LO}}$ the (N)LO correction to the inclusive jet
cross-section in the $p_t$ bin under consideration.
These expressions also require the inclusive jet cross-section, both
at LO and NLO, to be split in flavour channels. This is done as for
the 3-jet LO distribution above using the flavour-aware NLOJet++
version used in~\cite{bsz-shapes}. Alternatively, we have also used
the ($R$-)matching scheme given by Eq.~(3.28) of~\cite{bsz-shapes}.

From Eq.~(\ref{eq:matched}) it is trivial to obtain differential
distributions in bins of $\rho$. Normalised distributions can then be
obtained by dividing the result by the NLO inclusive jet cross-section
$\sigma_{\text{jet,LO}}+\sigma_{\text{jet,NLO}}$.\footnote{Note that
  this normalisation procedure gives consistent results when computing
  the uncertainties on the matched distributions.}

The uncertainties on the distributions come from four sources:
renormalisation and factorisation scales, resummation uncertainty and
matching uncertainty. The first two are estimated using the 7-point
rule~\cite{7point-rule}. The resummation uncertainties are obtained by
varying $\rho$ in Eqs.~(\ref{eq:nll-resum}) and~(\ref{eq:radiator})
between $\rho/2$ and $\rho$, introducing the appropriate correction
--- $\pm \log(2)R'$ in the exponent in~(\ref{eq:nll-resum}) --- to
maintain NLL accuracy. The matching uncertainty is estimated by
considering both the log-$R$ and $R$ matching schemes. We take the
central value from the central scale choice and the uncertainty from
the envelope of the 11 scale variations.\footnote{Seven factorisation
  and renormalisation scales, two resummation scales and two matching
  schemes.}

\begin{figure*}
\includegraphics[width=0.48\textwidth,page=2]{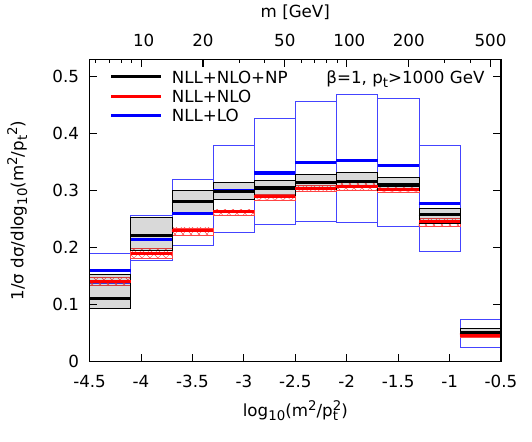}%
\hfill%
\includegraphics[width=0.48\textwidth,page=3]{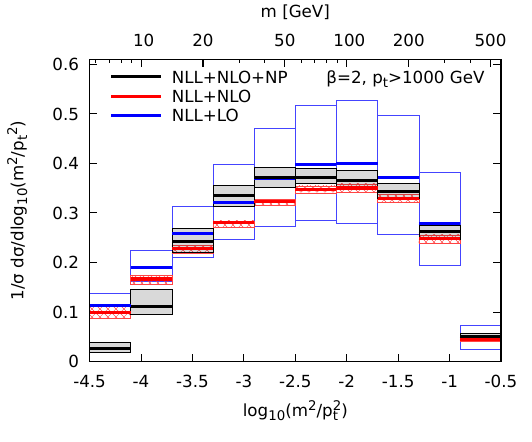}
\caption{Resummed and match theoretical predictions for the \SD jet
  mass distribution, for two different values of the angular exponent
  $\beta=1$ (left) and $\beta=2$ (right), $\zcut=0.1$ and $R=0.8$. The colours
  correspond to different accuracy of the calculation, as detailed in
  the legend.}
\label{fig:final-prediction}
\end{figure*}

%%%%%%%%%%%%%%%%%%%%%%%%%%%%%%%%%%%%%%%%%%%%%%%%%%%%%%%%%%%%%%%%%%%%%%%%%% 
%\section{Non-perturbative corrections}\label{sec:np}
\paragraph{Non-perturbative corrections.}

Power corrections induced by non-perturbative (NP) effects can be
estimated for \SD using the same approach as the equivalent
calculation for mMDT presented in Section~8.3.3 of Ref.~\cite{mMDT}.
We have to take into account two effects: (i) the mass of the SD jet
will be affected by NP corrections, (ii) NP effects can shift the
momentum of the subjets and alter the SD condition.

First, the mass shift can be written as (see~\cite{Dasgupta:2007wa})
$\delta m^2 = C_R\Lhadr p_t R_{\text{eff}}$, where $R_{\text{eff}}$ is
the effective jet radius after grooming, i.e. for a mass $m$ and
subjets passing the \SD condition with a momentum fraction $z$,
$R_{\text{eff}}=m/(p_t\sqrt{z(1-z)})$.
Following the same steps as in Ref.~\cite{mMDT} we
obtain\footnote{Although, instead of averaging $R_{\text{eff}}$ over
  $z$, we have kept explicit the $z$ dependence of $R_{\text{eff}}$
  and averaged the final correction over $z$.}
\begin{equation}\label{eq:hadr-mshift}
\frac{d\sigma}{dm}\bigg|_{\text{hadr}}^{(m\text{ shift})}
 = \frac{d\sigma}{dm}\bigg|_{\text{pert}}
\bigg(1+\frac{C_R\Lhadr}{m}\:\frac{\zsd^{-1/2}-\Delta_i}{\Lsd+B_i}\bigg),
\end{equation}
with $\zsd=\zcut^{\frac{2}{2+\beta}}\big(\frac{m}{p_tR}\big)^{\frac{2\beta}{2+\beta}}$,
$\Lsd=\log(1/\zsd)$ and
\begin{equation}
  \Delta_q=\frac{3\pi}{8}\quad\text{ and }\quad\Delta_q=\frac{(15C_A-6n_fT_R)\pi}{32\,C_A}.
\end{equation}

Then, hadronisation will shift the momentum of the softer subjet by an
average $\delta p_t=-C_A\Lhadr/R_{\text{eff}}$, where we have taken
into account that the softer subjet typically corresponds to a gluon
emission.
This means that emissions which were perturbatively passing the \SD
condition, with $\zsd<z<\zsd-\delta p_t/p_t$, will fail the \SD
condition after hadronisation, leading to a reduction of the
cross-section
\begin{equation}\label{eq:hadr-ptshift}
\frac{d\sigma}{dm}\bigg|_{\text{hadr}}^{(p_t\text{ shift})}
 = \frac{d\sigma}{dm}\bigg|_{\text{pert}}
\bigg(1-\frac{C_A\Lhadr}{m}\:\frac{\zsd^{-1/2}}{\Lsd+B_i}\bigg).
\end{equation}
The final hadronisation correction includes
both~(\ref{eq:hadr-mshift}) and~(\ref{eq:hadr-ptshift}). Both terms
are proportional to
$\frac{\Lambda_{\text{hadr}}}{p_t}\big(\frac{p_t}{m}\big)^{\frac{2+2\beta}{2+\beta}}$,
which increases with $\beta$ and has the appropriate limits for
$\beta\to\infty$ and $\beta\to 0$. 

A similar calculation can be carried out for the Underlying Event
(UE) contamination. In this case we have $\delta
p_t=\Lambda_{\text{UE}}\pi R_{\text{eff}}^2$ and $\delta
m^2=\frac{1}{2}\Lambda_{\text{UE}}p_tR_{\text{eff}}^4$. Following the
same steps as above, we find
\begin{equation}\label{eq:UE-mshift}
  \frac{d\sigma}{dm}\bigg|_{\text{UE}}
 = \frac{d\sigma}{dm}\bigg|_{\text{pert}}
\bigg(1+\frac{\Lambda_{\text{UE}} m^2}{p_t^3R^3}\:\frac{\zsd^{-2}(1-f_{m,i})}{\Lsd+B_i}\bigg),
\end{equation}
where the $1$ in the numerator corresponds to the $p_t$ shift and the
$f_{m,i}$ term corresponds to mass-shift effects, with
\begin{align}
f_{m,q} &= \frac{1+3\zsd+2\zsd^2(3\Lsd-2)}{4},\\
f_{m,g} &= \frac{1+2\zsd+3\zsd^2(2\Lsd-1)}{4} + \frac{n_fT_R}{C_A}\zsd(1-\zsd).\nonumber
\end{align}
This time, both sources of corrections give an effect proportional to
$\frac{\Lambda_{\text{UE}}}{p_t}\big(\frac{p_t}{m}\big)^{\frac{2\beta-4}{2+\beta}}$,
which increase with $\beta$ and has the expected
$\Lambda_{\text{UE}}p_t/m^2$ behaviour in the limit
$\beta\to \infty$.

In Fig.~\ref{fig:np-ratio}, we compare our analytic findings (dashed
lines) with the Monte-Carlo simulations, obtained with \pythia \
8.223 \cite{pythia8} (Monash 13~\cite{tuneM13} tune, solid
lines).
We consider both hadronisation corrections (left) and UE
effects (right), for a range of $\beta$ values.
UE effects are seen to be much smaller than hadronisation
corrections.
In the region where $\Lambda_{\text{hadr,UE}}\ll m \ll p_t$, our
analytic calculation captures the main features observed in
the simulations, including the increase with $\beta$ and the
global trend in $\rho$. 
At smaller mass, Pythia simulations exhibit a
peak in the hadronisation corrections which is beyond the scope of our
power-correction calculation. 

Even if the above analytic approach to estimating NP effects is
helpful for a qualitative understanding, it is unclear how
reliable it would be for phenomenology.
For example, hadron masses, which are neglected here, would have an
additional effect, even at large mass. Thus, the analytic estimates
can, at best, be trusted up to a fudge factor and analytic results can
not be trusted at small mass (see also~\cite{Dasgupta:2016bnd}).

As for our mMDT calculation~\cite{our_mmdt}, for our final
predictions, we have therefore decided to estimate NP corrections
using different Monte-Carlo simulations: \herwig~6.521~\cite{herwig}
with the tune AUET2~\cite{tuneAUET2}, \pythia~6.428~\cite{pythia6}
with the Z2~\cite{tuneZ2} and Perugia~2011~\cite{tuneP2011} tunes, and
\pythia \ 8.223~\cite{pythia8} \ with the 4C~\cite{tune4C} and
Monash~13~\cite{tuneM13} tunes.
For each Monte-Carlo, we compute the ratio between the full simulation
and the parton level. The average result is taken as the average NP
correction, and the envelope as the uncertainty which is added in
quadrature to the perturbative uncertainty.

%%%%%%%%%%%%%%%%%%%%%%%%%%%%%%%%%%%%%%%%%%%%%%%%%%%%%%%%%%%%%%%%%%%%%%%%%%
%\section{Final predictions and Conclusions}\label{sec:final}
\paragraph{Final predictions and conclusions.}

Our final predictions, are presented for $\beta=1$ (left) and
$\beta=2$ (right) in Fig~\ref{fig:final-prediction}.
To highlight our key observations, we present our final results at NLL
matched to NLO and including NP corrections (labelled NLL+NLO+NP), as
well as pure perturbative results (NLL+NLO) and results obtained when
matching to LO only (NLL+LO).
The most striking feature that we observe is that matching to NLO not
only affects quite significantly the central value of our prediction,
but also significantly reduces the uncertainty
% (dominated by normalisation uncertainty at LO)
across the entire spectrum.

Then, we see that NP corrections remain small over a large part of the
spectrum, although they start being sizeable at larger mass when the
angular exponent $\beta$ increases.
The fact that \SD observables can be computed precisely in
perturbative QCD, with small NP corrections, makes them interesting for
further phenomenological studies, including other observables like
angularities or attempts to extract the strong coupling constant from
fits to the data. 

Finally, we note that these predictions have recently been compared to
experimental results obtained by the ATLAS collaboration in
Ref.~\cite{atlas-sd}.
A good overall agreement between data and theory is observed,
especially in the region where NP corrections are small.

\begin{acknowledgements}
We thank Ben Nachman and Gavin Salam for many useful discussions.
SM and LS would like to thank the IPhT Saclay for hospitality during
the course of this project.
GS's work is supported in part by the French Agence Nationale de la
Recherche, under grant ANR-15-CE31-0016 and by the ERC Advanced Grant
Higgs@LHC (No.\ 321133).
\end{acknowledgements}

\end{document}